\documentstyle[11pt,newpasp,twoside]{article}
\def\edcomment#1{\iffalse\marginpar{\raggedright\sl#1\/}\else\relax\fi}
\reversemarginpar

\begin{document}
\title{Magnetically Driven Warping and Precession of Accretion Disks:
Implications for ``Exotic'' Stellar Variabilities}
 \author{Dong Lai}
\affil{Department of Astronomy, Cornell University, Ithaca, NY 14853, USA}

\begin{abstract}
The inner region of the accretion disk around a magnetized star is
subjected to magnetic torques that induce warping and precession of the
disk. These torques arise from interactions between the stellar field and
the induced electric currents in the disk.
These novel magnetic effects give rise to some
``exotic'' stellar variabilities, and may play an important role
in explaining a number of puzzling behaviors related to disk accretion 
onto magnetic stars, such as mHz QPOs in X-ray pulsars,
long-term periodicities of X-ray binaries (including precession of jets),
low-Frequency (10-50~Hz) QPO's in low-mass X-ray binaries, and 
photometric variabilities of T Tauri stars.
\end{abstract}

\section{Introduction}

Disk accretion onto a magnetic star occurs in a variety of astrophysical 
contexts, including accreting neutron stars, white dwarfs and 
pre-main-sequence stars (e.g., Frank et al.~1992). The basic
picture of disk--magnetosphere interaction is well-known: at large radii the
disk is unaffected by the stellar magnetic field; a somewhat sudden
transition occurs when the stellar field disrupts the disk at the
magnetospheric boundary, and channels the plasma onto the polar caps of the
star. The magnetosphere boundary is located where 
the magnetic and plasma stresses balance, 
\begin{equation}
r_m=\eta\,\mu^{4/7}(GM\dot M^2)^{-1/7},
\end{equation}
where $M$ and $\mu$ are the mass and magnetic moment of the central star,
$\dot M$ is the mass accretion rate, $\eta$ is a dimensionless constant
of order unity.

Because of its intrinsic importance for a wide range of astrophysical 
systems, a large number of theoretical papers have been written on the 
subject of the interaction between accretion disks and magnetized stars
(see references in Lai 1999 and in Shirakawa \& Lai 2002a,b), 
and numerical study of this problem
is still in its infancy. Outstanding issues remain, including the 
efficiency of field dissipation in/outside the disk, whether the disk excludes
the stellar field by diamagnetic currents or the field can penetrate a large 
fraction of the disk, whether the threaded field remains closed (connecting the
star and the disk) or becomes open by differential shearing, and whether/how
magnetically driven wind is launched from the disk or the
magnetosphere/corotation boundary. 
 
Many previous theoretical papers have, for simplicity, adopted
the idealied geometry in which the magnetic axis, the spin axis and 
the disk angular momentum are aligned. However, in Lai (1999), it was 
shown that under quite general conditions, the stellar magnetic
field can induce warping in the inner disk and make the disk 
precess around the spin axis (see \S 2). Such magnetically driven warping 
and precession open up new possibilities for the dynamical behaviors of 
disk accretion onto magnetic stars, and may explain some of the observed
variabilities in different stars (including compact objects).  

\def\be{\begin{equation}}
\def\ee{\end{equation}}
\def\ba{\begin{eqnarray}}
\def\ea{\end{eqnarray}}
\def\go{\mathrel{\raise.3ex\hbox{$>$}\mkern-14mu
             \lower0.6ex\hbox{$\sim$}}}
\def\lo{\mathrel{\raise.3ex\hbox{$<$}\mkern-14mu
             \lower0.6ex\hbox{$\sim$}}}
\def\cJ{{\cal J}}
\def\cQ{{\cal Q}}
\def\cO{{\cal O}}

\section{Magnetically Driven Warping/Precession}

Lai (1999) shows that the inner region of the accretion disk onto a rotating
magnetized central star is subjected to magnetic torques which induce 
warping and precession of the disk. The origin of these torques lies in
These magnetic torques result from the interactions between
the accretion disk and the stellar magnetic field.
Depending on how the disk responds to the stellar field,
two different kinds of torque arise:
(i) If the vertical stellar magnetic field $B_z$ penetrates the disk,
it gets twisted by the disk rotation
to produce an azimuthal field $\Delta B_\phi=\mp\zeta B_z$ that has different
signs above and below the disk ($\zeta$ is the azimuthal pitch of the field
line and depends on the dissipation in the disk), and a radial surface current
$K_r$ results. The interaction between $K_r$ and the stellar $B_\phi$ gives
rise to a vertical force. While the mean force (averaging over the azimuthal
direction) is zero, the uneven distribution of the force induces a net
{\it warping torque} which tends to misalign the angular momentum of the disk
with the stellar spin axis.
(ii) If the disk does not allow the vertical stellar field
(e.g., the rapidly varying component of $B_z$ due to stellar rotation)
to penetrate, an azimuthal screening current $K_\phi$ will be induced on the
disk. This $K_\phi$ interacts with the radial magnetic field $B_r$
and produces a vertical force. The resulting {\it precessional torque}
tends to drive the disk into retrograde precession around the stellar spin
axis.

In general, both the magnetic warping torque and the precessional torque are
present. For small disk tilt angle $\beta$ (the angle between the disk normal
and the spin axis), the precession angular frequency and warping rate
at radius $r$ are given by 
\ba
&&\Omega_p (r)=\frac{\mu^2}{\pi^2 r^7\Omega(r)\Sigma(r) D(r)}F(\theta),
\label{eqn:Omega_p}\\
&&\Gamma_w (r)=\frac{\zeta\mu^2}{4\pi r^7\Omega(r)\Sigma(r)}\cos^2\theta,
\label{eqn:Gamma_w}
\ea
where $\mu$ is the stellar magnetic dipole moment, $\theta$ is
the angle between the magnetic dipole axis and the spin axis,
$\Omega(r)$ is the orbital angular frequency, and $\Sigma(r)$ is the surface
density of the disk. [Note that the stellar spin
frequency $\Omega_s$ does not appear in eqs.~(2) \& (3) 
since the variation of the field geometry due to the spin has been averaged
out; this is justified because $\Omega_s\gg |\Omega_p|,~|\Gamma_w|$.]
The dimensionless function $D(r)$ is given by
\be
D(r)={\rm max}~\left(\sqrt{r^2/r^2_{\rm in}-1}, \sqrt{2H(r)/r_{\rm in}}\right)
\label{eqn:D(r)},
\ee
where $H(r)$ is the half-thickness and $r_{\rm in}$ is the inner radius of the
disk. The function $F(\theta)$ depends on the dielectric property of the
disk. We can write
\be
F(\theta)=2f\cos^2\theta-\sin^2\theta,
\ee
so that $F(\theta)=-\sin^2\theta$ if only the spin-variable vertical field is
screened out by the disk ($f=0$), and $F(\theta)=3\cos^2\theta-1$ if all
vertical field is screened out ($f=1$). In reality, $f$ lies between 0 and 1.
For concreteness, we shall set $F(\theta)=-\sin^2\theta$ in the following.

We also note the effect of {\it magnetically driven resonances}. 
For a general magnetic field--disk geometry,
the vertical magnetic force on a disk element varies
with the stellar rotation period. This gives rise to a number of 
{\it vertical resonances} in the disk. Similarly. there exist 
{\it epicyclic resonances} due to the time-dependent radial magnetic
force. Although the force expressions are model-dependent,
the existence of the resonances appears to be inevitable.
These magnetically driven resonances are somewhat similar to the corotation
resonance and Lindblad resonances in gravitational systems. 
The resonances may act as an extra source (in addition to the non-resonant
precessional and warping torques discussed above) for generating bending 
waves and spiral waves in the disk. Near the resonances, fluid elements 
undergo large out-of-plane and radial excursions, which may lead to
thickening of the disk. This may be analogous to the 
Lorentz resonances (which occur when charged particles
move around a rotating magnetic field) in the jovian ring
(e.g., Schaffer \& Burns 1992). However, because of the
fluid nature of the disk, the resonances may not lead to sharp edges 
in the disk.

\section{Dynamics of Warped Disks, Effects of Viscosity, Global Warping Modes
and Nonlinear Evolution}

\subsection{Effects of Viscosity}

Since the magnetic torque drives the disk tilt,
while the viscosity reduces the tilt, one can derive the 
criterion for the warping instability. Roughly speaking, the disk warp
can grow if the timescale associated with the warping torque is shorter than
the viscous time $r^2/\nu_2$ [where
$\nu_2$ is the viscosity (measuring the $r$-$z$ stress)
associated with reducing disk tilt].
Since the warping torque is a steep function
of $r$, the warping instability occurs only inside
a critical radius $r_w$. Our analysis (Lai 1999) shows that 
local warping torque can overcome viscous damping when 
\be
\Gamma_w >{2\pi^2}{\nu_2\over r^2}\Longleftrightarrow
{\rm Instability}.
\label{criterion}\ee
Assuming that $\nu_2/\nu_1$ is independent of $r$, the above equation reduces
to
\be
r<r_w=\left({3\,\zeta\cos^2\theta\over 8\pi^2\cJ}{\nu_1\over\nu_2}
\right)^{2/7}\left({\mu^4\over GM\dot M^2}\right)^{1/7},
\ee
where we have used
$\Sigma=({\dot M/3\pi\nu_1}){\cal J}$, 
and ${\cal J}$ is a dimensionless function of $r$ which approaches unity in the
region far from the inner edge of the disk.
Thus $r_w$ is typically a few times the canonical Alf\'ven radius
(the magnetosphere boundary). Therefore, 
as the disk approaches the magnetosphere,
its normal vector $\hat l$ will tend to be tilted with respect to the stellar
spin even if at large radii it is aligned with the spin axis. 

Another aspect of the viscous effect is what we call 
``{\it Magnetic Bardeen-Petterson Effect}''.
Because of the magnetic precessional torque,
the tilted disk will be driven into differential precession (with the
precession rate dependent on $r$). By analogy with the
well-known Bardeen-Petterson effect (i.e., the inner region of an accretion 
disk undergoing Lense-Thirring precession around a rotating black hole
tends to align itself with the equatorial plane of the black hole; 
see Bardeen \& Petterson 1975),
we expect that the magnetically driven
precession also tends to damp the tilt of the inner disk through the action of
viscosity. Setting $\Omega_p$ equal to $\nu_2/r^2$, we obtain 
the magnetic Bardeen-Petterson radius:
\be
r_{\rm MBP}=\left({3\sin^2\theta\over\pi\cJ D}{\nu_1\over\nu_2}\right)^{2/7}
\left({\mu^4\over GM\dot M^2}\right)^{1/7}.
\ee
Inside $r_{\rm MBP}$, the combined effect of viscosity and precession
tends to align the disk normal with the spin axis. 
We see that typically $r_{\rm MBP}$ is of the same order as $r_w$ (the
warping radius) and $r_m$
(the magnetosphere radius). Thus the precessional torque 
has an opposite effect on the
disk tilt as the warping torque. However, 
because of the broad warp-alignment transition expected for the magnetic
Bardeen-Petterson effect and the long timescale involved, we expect that the 
precession-induced alignment will be overwhelmed by the warping instability.

\subsection{Global Warping Modes and Nonlinear Evolution}

Since the precession rate $\Omega_p(r)$ depends strongly on $r$,
coupling between different rings is needed to produce a global coherent
precession. The coupling can be achieved either by viscous stress or
through bending waves (e.g., Papaloizou\& Pringle~1983; Papaloizou \&
Terquem~1995). In the viscosity dominated regime
(i.e., the dimensionless viscosity parameter $\alpha$ greater than $H/r$),
the dynamics of the warps can be studied using the formalism of
Papaloizou \& Pringle (1983) (see also Ogilvie~1999;
Ogilvie \& Dubus~2001). We model the disk as a collection of
rings which interact with each other via viscous stresses.
Each ring at radius $r$ has the unit normal vector ${\bf\hat l}(r,t)$.
In the Cartesian coordinates, with the $z$-axis along the neutron star spin,
we write
${\hat{\bf l}}=(\sin\beta\cos\gamma,\sin\beta\sin\gamma,\cos\beta)$,
with $\beta(r,t)$ the tilt angle and $\gamma(r,t)$ the twist angle.
For $\beta\ll 1$, the dynamical warp equation for ${\hat{\bf l}}$
(Lai 1999; see Papaloizou \& Pringle 1983; Pringle 1992) reduces to
an equation for $W(r,t)\equiv \beta(r,t)e^{i\gamma(r,t)}$:
\ba
&&\frac{\partial W}{\partial t}-
\left[\frac{3\nu_2}{4r}\left(1+\frac{2r\cJ'}{3\cJ}\right)
+\frac{3\nu_1}{2r}(\cJ^{-1}-1)\right]
\frac{\partial W}{\partial r}\nonumber\\
&&\qquad\qquad =\frac{1}{2}\nu_{2}\frac{\partial^2 W}{\partial r^2}
+i\Omega_pW+\Gamma_wW,\label{eqn:evolution}
\ea
where $\cJ'=d\cJ/dr$ (we assume that the ratio of $\nu_2$ to $\nu_1$ is
constant). In deriving the above equation, we have used the relations for the
radial velocity and surface density: $v_r=-3\nu_1\cJ^{-1}/2r$ and
$\Sigma={\dot M}\cJ/3\pi\nu_1$.
The values and functional forms of $\nu_1$, $\nu_2$, $\Omega_p$,
$\Gamma_w$ and the dimensionless function $\cJ(r)$ depend on disk models
(see Shirakawa \& Lai 2002a,b for details).

Shirakawa \& Lai (2002) carried out a global analysis
of warping/precession modes in a viscous accretion disk,
and show that under a wide range of conditions, the magnetic warping torque 
can overcome viscous damping and make the mode grow. The warping/precession
modes are concentrated near the inner edge of the disk (at the
magnetosphere-disk boundary), and can give rise to variabilities or
quasi-periodic oscillations (QPOs) in the X-ray/UV/optical fluxes from X-ray
pulsars (see \S 4). Pfeiffer \& Lai (2003) studied the nonlinear evolution
of the warping-precession modes, and found that the mode tends to saturate
at a large amplitude (its value depends on the parameters of the system).
The implications of such nonlinear behavior remain to be understood.

\section{Applications}

The magnetically driven warping instability and precession 
help explaining a number of observational puzzles related to 
stellar variabilities (for more details, see Lai 1999; 
Shirakawa \& Lai 2002ab; Pfeiffer \& Lai 2003).

\smallskip
{\bf (i) Milli-Hertz QPO's in Accreting X-ray Pulsars}:
Quasi-Periodic Oscillations (QPO's)  with frequencies $1-100$~mHz have been
detected in at least 11 accreting X-ray pulsars.
These mHz QPOs are often interpreted in terms of the
beat frequency model (BFM), in which the observed QPO frequency represents the
beat between the Keplerian frequency $\nu_K$ at the inner disk radius
and the NS spin frequency $\nu_s$, 
or in terms of the Keplerian frequency model (KFM), in which
the QPOs arise from the modulation of the X-rays by some inhomogeneities in the
inner disk at the Keplerian frequency. However for several sources, more than
one QPOs have been detected and the difference in the QPO frequencies is not
equal to the spin frequency. Thus KFM and/or the BFM cannot be the whole story.
Also note that in both the KFM and the BFM, it is always postulated that the
inner disk contain some blobs or inhomogeneities, whose physical origin is
unclear.
In Shirakawa \& Lai (2002) we suggest a ``Magnetic Disk Precession Model''
for the mHz variabilities and QPOs of accreting X-ray pulsars.
The magnetically driven precession of the warped inner disk (outside
but close to the magnetosphere boundary) can modulate
X-ray/UV/optical flux in several ways. We identify $\nu_{\rm QPO}$
with the global precession frequency driven by the magnetic
torques. Our calculations show that under a wide range of conditions,
the warping/precession mode is concentrated near the disk inner edge, and
the global mode frequency is equal to $A=0.3-0.85$ (depending on details of
the disk structure) times the magnetically driven precession frequency at
$r_{\rm in}=r_m$. An examination of the observed properties
of mHz QPOs in several systems (such as 4U 1626-67) suggests that
some hitherto unexplained QPOs are likely to be results of magnetically
driven disk warping/precession (see Chakrabarty et al.~2001).

\smallskip
{\bf (ii) Spin evolution of accreting X-ray pulsars}: Recent long-term,
continuous monitoring of X-ray pulsars with the BATSE instrument on the Compton
Gamma Ray Observatory (CGRO) has revealed a number of puzzling behaviors of the
spins of these objects (see Bildsten et al.~1997 and references therein). 
Several well-measured disk-fed systems (e.g., Cen X-3, GX 1+4 and 4U 1626-67) 
display sudden transitions between episodes of steady
spin-up and spin-down, with the absolute values of spin torques approximately
equal (to within a factor of a few). The transition timescale
ranges from days to years. It is likely that the  magnetically driven disk
warping may be a crucial ingredient in determining the spin behaviors of
accreting X-ray pulsars. With the magnetic warping torque, the perpendicular
state is an ``attractor''. The observed sign switching of $\dot\omega_s$ (spin
derivative) in several X-ray pulsars may be associated with the ``wandering''
of the inner disk around this ``attractor''. Rough estimate based on the
magnetic torque indicates that the switching timescale (which depends
on the stellar field strength, the disk parameters and geometry)
ranges from days to years, in agreement with observations. 

\smallskip
{\bf (iii) Quasi-periodic oscillations in low-mass X-ray binaries}: 
Rapid variability in low-mass X-ray binaries, containing
weakly magnetized ($B\sim 10^8$~G) neutron stars,
has been studied since the discovery of the so-called 
horizontal-branch oscillations (HBOs) (see van der Klis 1998).
The HBOs are quasi-periodic oscillations
(QPOs) (with $Q$-value $\nu/\Delta\nu$ of order a few)
which manifest as broad Lorentzian peaks in the X-ray power spectra 
with centroid frequencies in the range of 15--60~Hz which are positively
correlated with the inferred mass accretion rate. Stella and Vietri (1998)
suggested that HBOs (and other low-frequency QPOs)
are associated with Lense-Thirring precession of the
inner accretion disk around the rotating NS. 
For this interpretation to be viable, the inner 
disk must be tilted with respect to the stellar spin axis. Since the
Bardeen-Petterson effect tends to keep the inner region of the disk 
(typically within 100--1000 Schwarzschild radii) co-planar with the star 
and radiation-driven warping is only effective at large disk radii
(Pringle 1996), another mechanism to drive warping in the inner disk 
is needed. The magnetic warping torque provides
a natural source for inducing disk tilt. Moreover, 
the magnetically driven (retrograde) precession 
rate is not negligible compared to the Lense-Thirring precession rate,
and will contribute to the total precession (Shirakawa \& Lai 2002a).

Other possible applications include:

{\bf (iv) Long-term (super-orbital) variabilities in X-ray binaries
(including precession of jets):} The well-known examples include
Her X-1 (35 days), LMC X-4 (30.4 days) and SS433 (164 days). It has always
been thought that these super-orbital periods are caused by 
precession of accretion disks, perhaps driven by the binary companion. 
However, the tidal torque from the companion is relevant only if the disk is
warped. Without any extra driver for the disk warp, the disk would be
flat. Magnetic field may play a role here.

{\bf (v) Photometric period variations of T Tauri stars:} 
T Tauri stars have magnetic fields of order 1~kG. Being magnetic,
they are variable. Most of the variabilities can be explained by rotating
cold spots or hot spots on the stellar surface. However, some of the
variabilities in classical T Tauri stars are not easy to understand
in this picture. For example, AA Tauri shows photometric variability
(by 1 mag) in different bands on timescales of 8.5~days, but there is
no clear color variation (see Bouvier et al.~1999). 
This and some other features can be naturally
explained by a warped inner disk which causes occultation of the photosphere
(see Carpenter et al.~2001 for possibly other examples).

\smallskip
I thank Akiko Shirakawa and Harald Pfeiffer for their important 
contributions. This work has been supported in part 
by NSF AST 9986740 and NASA 
NAG 5-8484, and by the Alfred P. Sloan foundation.

\end{document}